\documentclass[
aps,%
12pt,%
final,%
notitlepage,%
oneside,%
onecolumn,%
nobibnotes,%
nofootinbib,%
superscriptaddress,%
noshowpacs,%
centertags]%
{revtex4}

\begin{document}


\title{THE REGION OF TRIGGERED STAR FORMATION W40: OBSERVATIONS AND MODEL}

\author{\firstname{L.~E.}~\surname{Pirogov}}
\email{pirogov@appl.sci-nnov.ru}
\affiliation{%
Institute of Applied Physics Russian Academy of Sciences, Nizhni Novgorod, Russia
}
\affiliation{%
Lobachevsky State University of Nizhni Novgorod, Nizhni Novgorod, Russia
}


\begin{abstract}

A ``collect and collapse'' model of triggered star formation is used to estimate the parameters
of ring-like structure consisted of a sequence of low-mass clumps in the W40 region.
The model parameters are close to the observed ones if the density of the cloud in which the HII zone is expanding is fairly high ($\ga 10^5$~cm$^{-3}$)
and the luminosity of the driving source exceeds previous estimate.
Probable reasons for the scatter of the observed parameters of the clumps are discussed.

\end{abstract}

\maketitle

\section{Introduction}

Early stages of the star formation process are far from fully understood
in spite of increasing amount of observational data available.
This is true, in particular, for regions of high-mass star formation which are
more rare, more distant, and evolve more quickly than regions of low-mass star formation, spend considerable time of their evolution inside the parent cloud (e.g. \cite{ZinneckerYorke}).
As massive stars evolve, they affect parent cloud through their stellar winds,
massive outflows, strong UV-radiation, and expansion of their HII zones.
These factors change the physical conditions and chemical composition of the cloud
and the expanding of HII zones ``sweep out'' the gas towards the periphery causing compression
and possibly trigger the formation of a new generation of stars.

The W40 region (Sharpless 64) \cite{Westerhout, Sharpless} contains large blister-type HII zone that lies at the edge of an extended molecular cloud (TGU 279-P7 \cite{Dobashi}) in the Aquila Rift complex.
According to recent estimates the distance to the central sources of the HII zone is
 $\sim 500$~pc \cite{Shuping}.
This puts W40 among the nearest regions of high-mass star formation.
The difference between CO and hydrogen recombination line velocities indicates that the HII zone lies at the front edge of the cloud closer to the observer \cite{ZeilikLada}.
The HII region expands into molecular cloud to the west.
Molecular gas is observed is observed at two different velocities which probably correspond to gas located behind and in front of the HII zone \cite{Vallee}.
The W40 region was studied at various wavelengths from the radio to the X-ray
(see the review \cite{RodneyReipurth} and \cite{Rodriguez, Shuping, Kuhn, Maury, Pirogov, Mallick}).

Continuum observations of dust emission \cite{Maury, Pirogov} have shown that the dust to the west of the HII zone is concentrated in clumps forming a ring.
The morphology of ionized gas \cite{Mallick, Pirogov} shows that there is another
compact HII zone near the main one.
The expansion of this zone could lead to formation of the ring-like structure due to the ``collect and collapse'' mechanism.
This model was proposed in \cite{ElmegreenLada} to describe the process
of triggered high-mass star formation.
Although there are more than a few examples where this mechanism
could be taking place
(e.g. \cite{Deharveng03, Deharveng08, Deharveng10, Ohlendorf, Samal})
there have been few quantitative comparisons between observational data and the model.
This can be related to the fact that the model is fairly simple compared
to real objects with inhomogeneous density structure.
In the W40 region which is located much more closer than most of the HII zones
where the collect and collapse mechanism could take place
the observed ring-like structure is more distinct and compact and is
much closer to the probable driving source of the HII region.
We present here quantitative comparison of the physical parameters derived
from observations with the model predictions
in order to determine whether the collect and collapse
model can predict parameters of the structure found in W40.

\section{The ring of dust clumps}

A 1.2 mm dust continuum emission map of W40 taken from \cite{Pirogov}
is given in the figure.
The dust is concentrated in a chain of clumps forming a ring.
The data of \cite{Maury} show that
the western branch of the ring extends to the northeast,
making the ring-like structure more distinct.
Some of the clumps are associated with the Class~0 and Class~I sources \cite{Maury},
indicating that low-mass star formation has started in the ring.
The Near-IR sources \cite{Shuping} and 3.6 cm compact VLA radio sources \cite{Rodriguez}
are also shown in the figure (left panel).

Most of the radio and IR sources are grouped near the driving source of the main HII zone (IRS~1A~South),
which is located to the east of the ring and appears to be a massive O9.5 star \cite{Shuping, Mallick}.
The IRS~5 infrared source, which is a B1 star and the probable source
of the neighboring compact HII zone \cite{Mallick, Pirogov}, is also marked in the figure.
Since IRS~5 is located within the region bounded by the ring, but is shifted
relative to its geometrical center, it is probable that the ring formed due
to the expansion of the compact zone around this source.
The ring is apparently not oriented face-on, and its south-eastern part
is closer to observer \cite{Pirogov}.

The formation of the ring-like structure from a sequence of fragments (clumps)
suggests the action of collect and collapse mechanism of triggered star formation \cite{ElmegreenLada}.
This mechanism is treated analytically in \cite{Whitworth} and the results
of that paper have been confirmed by model calculations \cite{Dale}.
According to the collect and collapse mechanism when an HII region expands
into molecular cloud, a layer of enhanced density forms between the shock front
and the ionization front.
Due to instability, the layer splits into uniformly spaced clumps, which can be
the sites of formation of next generation of stars.
Such ring-like structures usually consisting of massive fragments ($M\sim 10-100~M_{\odot}$)
are observed at the edges of some HII zones (e.g. \cite{Deharveng03, Deharveng08, Ohlendorf}).
However, in the case of W40, the masses of clumps in the ring
($0.4-8$~$M_{\odot}$ \cite{Pirogov}) are considerably lower than is usually observed in such structures.

\section{Analytical model}

The evolution of the layer formed while an HII zone expands into a medium
consisted of neutral gas with uniform density is considered in \cite{Whitworth}.
Analytical expressions for the time at which fragmentation in the layer occurs ($t_{frag}$),
the layer radius ($R_{frag}$) and column density of the layer ($N_{frag}$) at this time,
the mean mass of  fragments ($M_{frag}$) and initial separation between fragments ($2r_{frag}$) have been derived.
They depend on the sound speed in the medium ($a$), on the density of atomic gas ($n$),
and on the luminosity of the central source of ionized radiation ($L$).
The expressions of \cite{Whitworth} are given below, where
$a_{0.2}=a/0.2$\,km/s, $L_{49}=L/10^{49}$\,photon/s, $n_3=n/10^3$\,cm$^{-3}$:

\begin{equation}
t_{frag}\sim 1.6\,{\text{Myr}}~~a_{0.2}^{7/11} L_{49}^{-1/11} n_3^{-5/11}~,
\label{eq1}
\end{equation}

\begin{equation}
R_{frag}\sim 5.8\,{\text{pc}}~~a_{0.2}^{4/11} L_{49}^{1/11} n_3^{-6/11}~,
\label{eq2}
\end{equation}

\begin{equation}
N_{frag}\sim 6\times 10^{21}\,{\text{cm}}^{-2}~~a_{0.2}^{4/11} L_{49}^{1/11} n_3^{5/11}~,
\label{eq3}
\end{equation}

\begin{equation}
M_{frag}\sim 23\,M_{\odot}~~a_{0.2}^{40/11} L_{49}^{-1/11} n_3^{-5/11}~,
\label{eq4}
\end{equation}

\begin{equation}
2r_{frag}\sim 0.8\,{\text{pc}}~~a_{0.2}^{18/11} L_{49}^{-1/11} n_3^{-5/11}~.
\label{eq5}
\end{equation}

Note that, when the gas is in molecular form, the atomic gas density
is twice of the molecular gas density \cite{Dale}.
We neglected this factor in our approximate estimates derived
from (\ref{eq1})--(\ref{eq5}).

It is clear from the above equations that, when the HII zone made by a star
emitting $10^{49}$~photons/s (main sequence O6 massive star \cite{Panagia})
expands into a cold cloud with moderate density ($T_{\rm KIN}\sim 10-15$~K, $n\sim 10^3$~cm$^{-3}$),
a layer of fragments with masses $\sim 25-40~M_{\odot}$ forms
at a distance $\sim 6$~pc after $\sim 1.6$~Myr.
The initial separation between centers of the fragments is $\sim 1$~pc.
Such fragments could become sites of formation of a new generation of stars,
including massive ones.
If an HII zone expands into a medium of higher density, the layer
of enhanced density will form on shorter scales and closer to the star.
This layer will split into fragments with lower sizes and masses.

The luminosity of IRS~5, the source that has probably formed the ring-like structure in W40,
is $\sim 2.4\times 10^{45}$~photons/s, which corresponds to a B1V star \cite{Mallick}.
IRS~5 is classified as a B1 star in \cite{Shuping}.
However, the luminosity could be appreciably underestimated,
as well as the luminosity of the main HII zone (see the discussion in \cite{Mallick}).
We used two luminosities in calculations: one close to the estimate of \cite{Mallick}
and the second an order of magnitude higher.
The results of comparing the model calculations data with the observed parameters
are given in the following section.

\section{Comparison of the model and the observed clump parameters}
\label{comparison}

The dense core of the molecular cloud near the HII zone in W40 has a structure
consisting of several components, including clumpy ring and gas which
that has not condensed into clumps \cite{Pirogov}.
The western clumps {\it 6--9} are associated with the N$_2$H$^+$ and NH$_3$
molecular line emission regions, while the CS emission is enhanced towards
the eastern clumps {\it 1--3} (Figure, right panel).
THe mean densities in clumps derived from continuum observations are:
$\sim 10^5-10^6$~cm$^{-3}$, while the densities calculated from molecular line
data are higher.
The gas densities of extended regions that do not correlate with the ring
and are observed mainly in the HCO$^+$(1--0) and HCN(1--0) lines (Figure, right panel)
should be also fairly high (the critical densities for the excitation these
transitions are $\sim 10^5-10^6$~cm$^{-3}$).
The ratios of the HCN(1--0) hyperfine components imply moderate optical depth \cite{Pirogov}.
The HCN(1--0) emission is apparently associated with regions of high density
and not with enhanced column density of the gas.
The spatial distributions of the different CO isotopic line intensities \cite{Zhu}
correlate weakly with the ring structure, and may be tracers of a more diffuse
envelope around the ring.
The clumpy ring probably formed due to the expansion of the HII zone associated
with IRS~5 into the parent dense cloud.
In this case the gas emitting in the HCO$^+$ and HCN lines may represent
material that has not experienced the expansion of the HII zone,
since it is more distant from the source.

The Table presents parameters calculated from (\ref{eq1})--(\ref{eq5})
for $a_{0.2}\approx 1.15$ ($T_{\rm KIN}=15$~K, $m=2.33~m_H$)
and for two luminosities, one close to the estimate of \cite{Mallick}
and the other an order of magnitude higher.
The density of the surrounding medium for which the model estimates are more
or less close to the observed parameters is given for each luminosity.
The kinetic temperature value was taken to be close to kinetic and/or
dust temperature estimates for the western clumps \cite{Pirogov, Maury}.
The eastern clumps {\it 1--3} associated with Class~I sources \cite{Maury}
have higher dust temperatures,
and are influenced by the ionization front of the main HII zone,
which probably changes their physical characteristics.
The ranges of physical parameters obtained from the observations \cite{Pirogov}
are given in the table for comparison.
The parameters of the eastern clumps {\it 1--3} are excluded from consideration.

The size of the H~II zone around IRS~5 has been approximately estimated in \cite{Mallick}
as the diameter of the ring of Class 0/I sources, $\sim 0.4$~pc.
However, this ring is probably not oriented face-on, with its eastern part
is closer to the observer and different distances from the probable
central source IRS~5 to the boundary of the ring in different directions
(see Figure).
The maximum distance between IRS~5 and the inner boundary of the ring
in the southern direction is $\sim 0.4$~pc
\cite{Pirogov}, which is twice the radius estimated in \cite{Mallick}.
The average of these two radius estimates, 0.3~pc, is given in the table
as an observed radius of the ring, $R_{frag}$.

To be compared with the time at which gas in the layer becomes gravitationally
unstable and splits into fragments ($t_{frag}$), the corresponding dynamical
age of the HII zone ($t_{dyn}$) is given in the table.
This was calculated from the standard expression for the time dependence
of the radius of an H~II zone \cite{Spitzer} for a given luminosity and density:
$R(t)=R_s(1+\frac{7\,c_{II}\,t}{4\,R_s})^{4/7}$,
where $R_s$ is the radius of the Str\"omgren zone which depends
on the source luminosity and the density of the medium (e.g. \cite{Mallick}),
and $c_{II}$ is the sound speed of the ionized gas, taken to be 11~km/s \cite{Mallick}.
The condition $t_{dyn}\ga t_{frag}$ is satisfied in the calculations.

The ranges of the peak column densities and masses calculated from
the dust continuum observations are given in the table as the observed values
of $N_{frag}$ and $M_{frag}$.
The standard gas-to-dust mass ratio, equal to 100 \cite{Pirogov} was adopted.
Variations of these values are more likely related to density variations
in the gas into which the HII zone is expanding.
Variations of the gas-to-dust mass ratio in clumps (Section~\ref{discussion})
could also be a possible reason.
The observed separations between fragments ($2r_{frag}$) lie in the range indicated in the table
due to probable density variations and projection effects.
It is possible that the separations between the western clumps {\it 7, 8}
and {\it 9} depend on these factors to a lesser extent.
The observed clump sizes (0.02--0.11~pc) and masses are typical for low-mass
star forming regions, but
there are several indications that the formation of stars with masses
higher than a solar mass
is occurring in the eastern part of the ring \cite{Pirogov}.

A comparison between the observed parameters of the clumpy ring and the
two sets of model parameters
(see the table) shows that the model with higher luminosity fits the observations better.
According to the model estimates, in both cases, the density of the medium
into which the HII zone expands considerably exceeds the value usually adopted
as a standard for neutral gas surrounding HII zones ($\sim 10^3$~см$^{-3}$).
A better agreement between the model estimates of the masses and
the separations of the fragmetns can be achieved if a lower temperature is adopted
for the medium ($\sim 10$~K).
The density of the medium is not appreciably changed in this case.
The small excess of the observed radius and column density of the layer
over the model estimates (see the table) could be connected with
the evolution of these parameters during the time that has passed
since the onset of fragmentation in the layer.

\section{Discussion}
\label{discussion}

The structure of the W40 dense core probably formed due to the influence
on the neutral material of two HII zones whose main driving sources
are IRS~1A~South and IRS~5 (Figure, left panel).
There is a cluster of sources near IRS~1A~South.
Two of these sources (IRS~2B, IRS~3A) are massive B4 and B3 stars,
respectively \cite{Shuping}.
The main H~II region to the west is bounded by dense molecular gas and is more extended in the eastern direction \cite{Mallick}.
Its emission overlaps with emission of several compact H~II zones (the figure, left panel; Fig.~10 from \cite{Mallick}).
The HII zone around IRS~5 is located to the northwest from the main zone,
has a more or less circular morphology, and is probably a distinct region.
The morphology of the ionized gas shows that, even taking into account possible projection effects,
it is not possible to explain the formation of the ring due to the influence
of the main HII zone, since its main driving source, IRS~1A, is located to the east of the ring.
Alternative mechanisms for triggered star formation such as, radiation-driven
implosion, are unlikely be able to explain the formation of the observed
ring-like structure with regularly spaced clumps.

As model calculations show (e.g. \cite{Bisbas}) radiation-driven implosion
can form globules of elongated (cometary or pillar-like) form that
are associated with already existing inhomogeneities in the medium.
Star formation occurs at their front edges (facing the star).
Moreover, this mechanism predicts the existence of velocity gradients inside the globules
\cite{LeflochLazareff}, which are not observed in the clumps forming the ring-like structure in W40.
Nevertheless, radiation-driven implosion could take place at the outer boundary
of the eastern branch of the ring, where the morphologies of the ionized gas
(according to observations with high spatial resolution), dust and dense molecular gas are similar.
The line-of-sight velocities of the gas associated with the eastern clumps
differ from the velocities of the remaining gas in the region, and this gas is contracting \cite{Pirogov}.
Therefore, the collect and collapse mechanism seems to be the most probable reason
for the formation of the ring if the luminosity of IRS~5 is higher than the estimate of \cite{Mallick}
and the medium where ionization front propagates has a high density.

The observed scatter of the clump masses, column densities, and separation
could be associated with density variations in the medium around IRS~5.
The scatter in the clump separations could be connected with both
density variations and projection effects.

Note that clump masses and column densities were calculated in \cite{Pirogov}
using dust continuum observations.
The gas-to-dust mass ratio was taken to be 100.
It cannot be ruled out that this ratio could vary in the dense gas near an HII zone.
For example, the HCN, HCO$^+$, CS and $^{13}$CO molecular emission is detected
inside the region bounded by the ring, where the dust emission is considerably
reduced \cite{Pirogov, Zhu, Maury}.
It is shown in \cite{Hopkins} that spatial fluctuations of the dust density
in turbulent clouds can occur independently of gas-density fluctuations
on scales compared with the sizes of protostellar cores.

It is possible the level of turbulence of the medium increases as shock propagates,
leading to spatial fluctuations in the gas-to-dust abundance ratio.
The observations show that the widths of the N$_2$H$^+$, NH$_3$ and H$^{13}$CO$^+$
lines associated with the western clumps are half the widths of the CS, HCN and HCO$^+$ lines
\cite{Pirogov}.
In addition to the optical depth effects
this could be related with different levels of turbulence in the gas
that emits effectively in these lines.
Thus, it is possible that the gas with higher level of turbulence emitting
in the CS, HCN and HCO$^+$ lines and spatially uncorrelated with the clumps
could have a lower dust abundance.
The hypothesis that there may be a relation between turbulence
and spatial fluctuations of dust abundance close to HII zones need to be studied further,
using both independent observational estimates of gas and dust abundances
and model calculations.

\section{Conclusion}

We have carried out
a comparison of the observed parameters of a ring-like structure consisting
of low-mass clumps in W40 and estimates produced using the collect and collapse model \cite{ElmegreenLada, Whitworth}.
The physical parameters of the clumps were taken from \cite{Pirogov}.
This comparison shows that parameters such as the radius of the ring,
the masses and gas column densities in the clumps, and the separation
between clumps are close to the model estimates if the density of the cloud
into which the HII zone expands is fairly high ($\ga 10^5$~см$^{-3}$)
and the luminosity of the source driving the HII zone exceeds the previous
estimate \cite{Mallick}.
Thus, W40 can be considered an example of the realization of the collect and collapse
mechanism at a relatively small distance from the source in a high-density medium.
The observed scatter of the physical parameters of the clumps could be
associated with density inhomogeneities in the medium where
ionization front propagates, and also with projection effects and turbulence
of the medium that results in fluctuations of the gas-to-dust mass ratio.

\begin{acknowledgments}
We thank D.~Z.~Wiebe for his question concerning quantitative estimates
of the parameters of the collect and collapse model for W40 which
inspired this study.
We also thank the referee for
valuable comments and questions that led to recalculation of the model parameters
and a significant makeover of the paper.
This work was partially supported by Russian Foundation for Basic Research
(grants 12-02-00861, 13-02-92697, 13-02-12220)
and the grant (the agreement of August 27, 2013 No. 02.В.49.21.0003
between The Ministry of education and science of the Russian Federation and Lobachevsky State University of Nizhni Novgorod).

\end{acknowledgments}

%
%

\newpage

\begin{figure}[t!]
\setcaptionmargin{5mm}
\onelinecaptionsfalse

\begin{minipage}[b]{0.485\textwidth}
 \includegraphics[width=\textwidth,angle=-90]{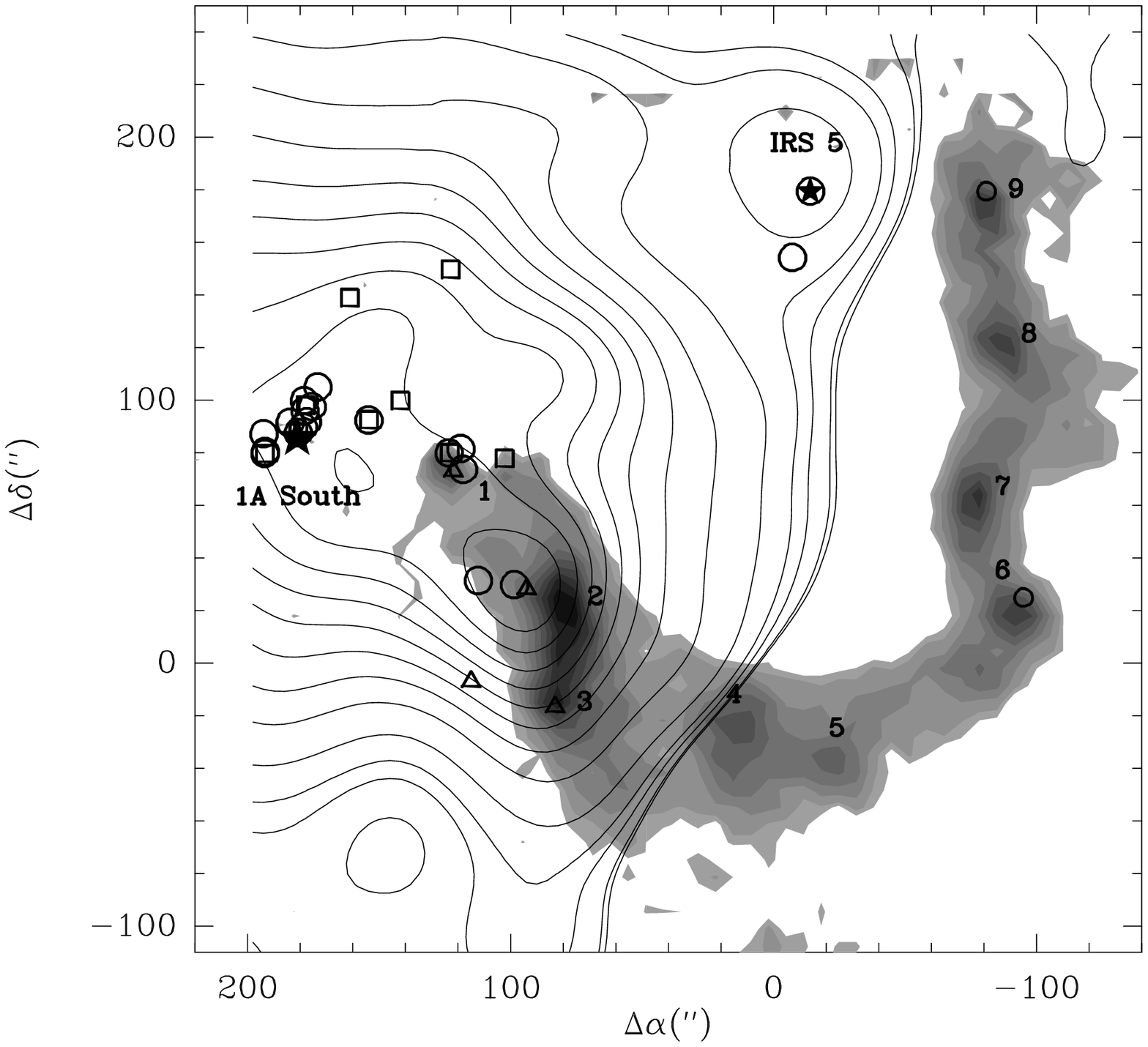}
\end{minipage}
\hspace{2mm}
\begin{minipage}[b]{0.485\textwidth}
 \includegraphics[width=\textwidth,angle=-90]{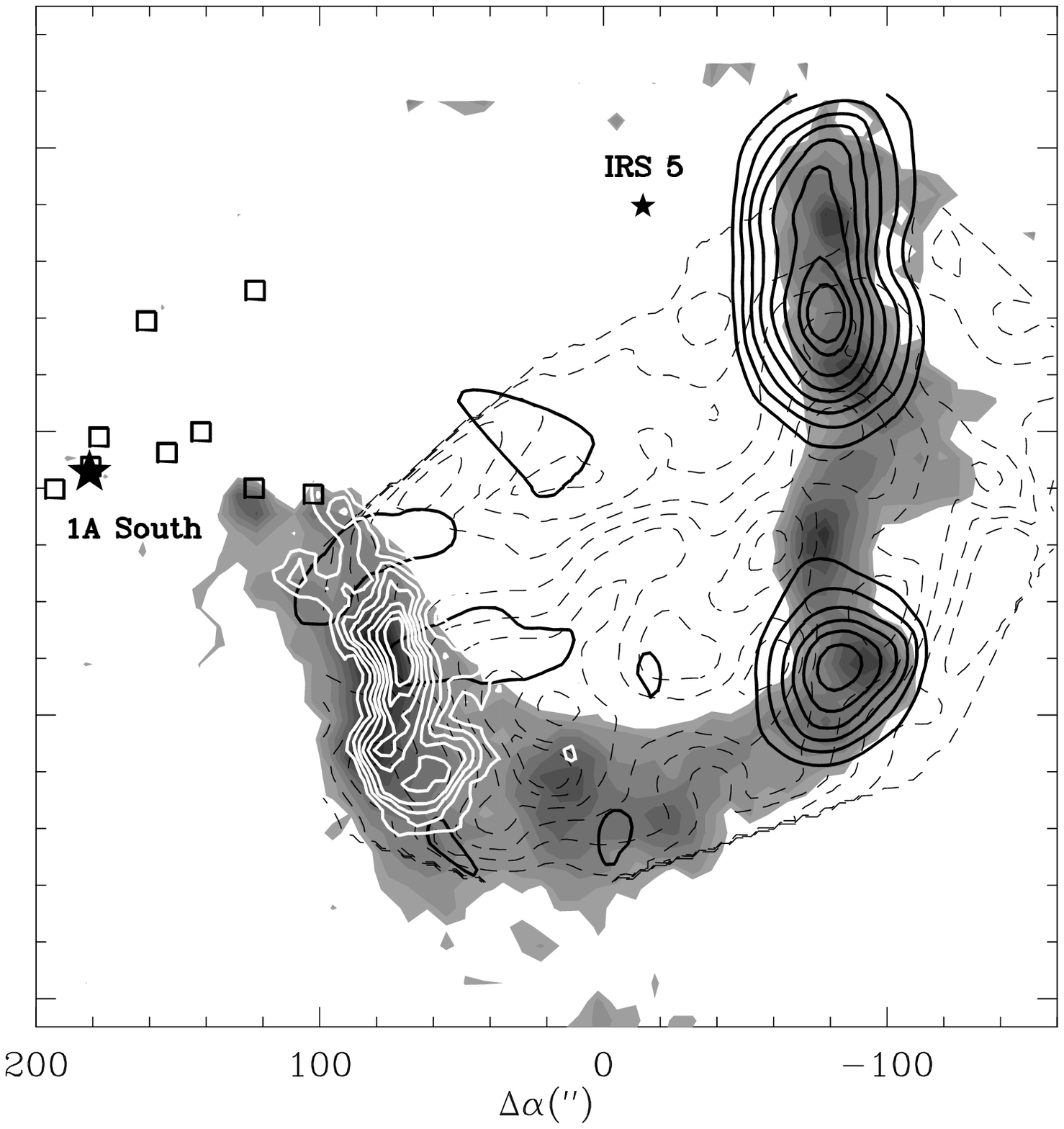}
\end{minipage}

\captionstyle{normal}
\caption{
Maps of dust, ionized and molecular gas emission in the W40 region
according to the data from \cite{Pirogov}.
The coordinates of the central position (the $^{13}$CO emission peak \cite{Zhu}) are:
R.A.(J2000)=18$^{h}$~31$^{m}$ 15.75$^{s}$,
Dec.(J2000)=--02$\degr$~06$'$~49.3$''$.
The clumpy ring is shown in greyscale.
Distinct clumps are denoted by numbers (left panel).
The spatial distribution of the ionized gas is shown as contours in the left panel.
The HCN(1--0) (dashed contours), NH$_3$(1,1) (dark contours) and CS(5--4) (light contours) molecular line maps are shown in the right panel.
Near IR-sources are shown as small squares \cite{Shuping}.
Compact VLA radio sources \cite{Rodriguez} are shown as larger circles in the left panel.
Class~0 and Class~I sources \cite{Maury} are marked by smaller circles and triangles, respectively
(left panel).
The driving source of the main HII zone, IRS~1A~South,
and the source of the separate H~II zone, IRS~5, \cite{Shuping} are shown
as stars.
}
\label{ring}
\end{figure}

\pagebreak

\begin{table}[p]
\setcaptionmargin{0mm}
\onelinecaptionsfalse
\captionstyle{flushleft}
\caption{Observed and model parameters of the clumps}
\begin{tabular}{l|c|c|c}
\hline
  Parameter &   Observed value  & \multicolumn{2}{c}{Model values}  \\
 \noalign{\smallskip}\cline{3-4}\noalign{\smallskip}
 &   \cite{Pirogov}              & $L (photons/s) =3\cdot 10^{45}$ & $L (photons/s) = 3\cdot 10^{46}$ \\
\noalign{\smallskip}\hline\noalign{\smallskip}
$n$(cm$^{-3}$)                           &                         &       $10^5$      &   $1.5\cdot 10^5$ \\
$t_{dyn}$ (Myr)                  &                          &        0.65        &    0.45   \\
$t_{frag}$ (Myr)                  &                          &       0.44        &     0.3  \\
$R_{frag}$ (pc)                           &  $\sim 0.3$       &        0.24          &     0.23\\
$N_{frag} (10^{22}$~cm$^{-2}$) & 4--11                &        2.4            &      3.6\\
$M_{frag} (M_{\odot}$)               &  2--6                 &        9.8            &     6.6\\
$2r_{frag}$ (pc)                           & $\sim 0.1-0.2$ &        0.27          &     0.18\\
\hline
\end{tabular}
\label{param}
\end{table}

\end{document}